\documentclass[12pt]{article}
 
\usepackage{epsfig}
\begin{document}


\baselineskip=14pt plus 0.2pt minus 0.2pt
\lineskip=14pt plus 0.2pt minus 0.2pt

\begin{flushright}
hep-th/0106233 \\
LA-UR-01-3190 \\
\end{flushright} 

\begin{center}
\large{\bf Existence of Bound States in Continuous 
$\mathbf{0<D<\infty}$
Dimensions} 
 
\vspace{0.25in}

\normalsize
\bigskip

Michael Martin Nieto\footnote{\noindent  Email:  
mmn@lanl.gov}\\
{\it Theoretical Division (MS-B285), Los Alamos National Laboratory\\
University of California\\
Los Alamos, New Mexico 87545, U.S.A. \\}

\normalsize

\vskip 20pt
\today

\vspace{0.3in}

{ABSTRACT}

\end{center}

\begin{quotation}

In modern fundamental theories there is consideration of higher
dimensions, often in the context of what can be written as a Schr\"odinger
equation. Thus, the energetics of bound states in different dimensions 
is of interest.  
By considering the quantum square well in continuous $D$ dimensions, it is
shown that there is always a bound state for $0<D \le 2$.  This binding is
complete for $D \rightarrow 0$ and exponentially small for 
$D \rightarrow 2_-$. For $D>2$, a finite-sized well is always needed for
there to be a bound state.  This size grows like $D^2$ as $D$ gets large. 
By adding the proper angular momentum tail a volcano, zero-energy, bound 
state can be obtained.

\vspace{0.25in}

\end{quotation}

\newpage

\baselineskip=.33in

\section{Introduction}

It is a well-known property of 1-dimensional quantum mechanics 
\cite{textbook2}-\cite{textbook4} that any (even infinitesimally small) 
potential well, that is bounded above by and asymptotes to zero energy,
supports a negative-energy bound state.  A proof by demonstration can be
given by noting that any finite well can be bounded from above by a square
well, and the square well can be explicitly solved to demonstrate a bound
state. 

A more sophisticated demonstration can use a Gaussian trial wave function to 
demonstrate, by the Rayleigh-Ritz variational method, that there must be a
bound state.  This particular  proof fails for higher dimensions.  
Since a square well of radius $a$ and depth $V_0$ in 3-dimensions 
must have $V_0a^2 > (\pi/2)^2$ for there to be a bound state, 
it can be tempting to state that this ends the discussion, 
but not quite \cite{martin}.   
[Except where noted, in this paper we use units $(\hbar^2/2m)=1$.]

What if the potential goes above zero?  If, 
in addition, it asymptotes above zero, then
one needs a finite-sized potential to have a bound state.  For example, 
the Morse and Rosen-Morse potentials  
\begin{equation}
V_{M}(z) = A_0\left[e^{-2z/a} - e^{-z/a}\right], ~~~~~~~~~
V_{RM}(z) = C_0 [1 + \tanh z/a] - \frac{U_0}{\cosh^2z/a} 
\end{equation} 
only have (negative-energy) bound states if 
\begin{equation}
A_0a^2 < 1/2, ~~~~~~~~~~~~~~~~~~
C_0a^2 < \left[1+2U_0a^2\right] - \left[1 +4U_0a^2\right]^{1/2}.
\end{equation}

\begin{figure}[h]
 \begin{center}
\noindent    
\psfig{figure=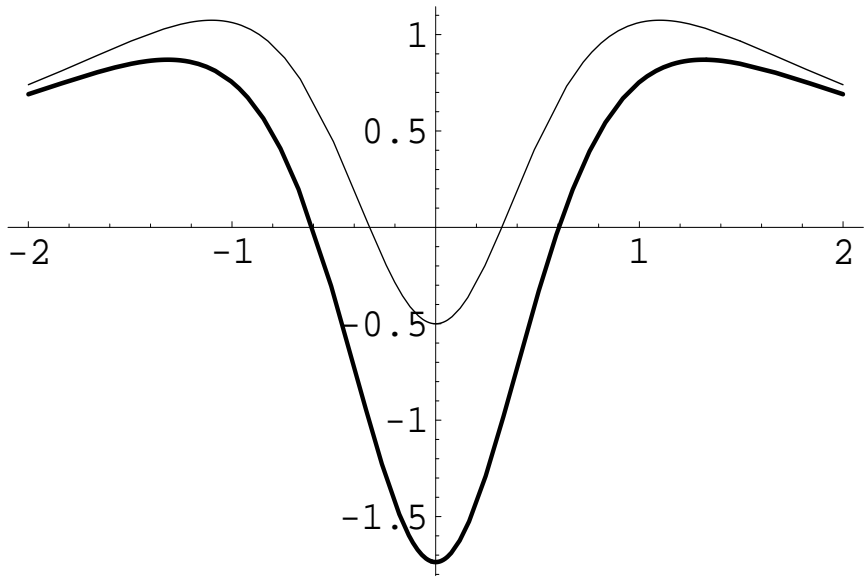,width=4.5in,height=3in}
\caption{The thin line shows the potential of Eq. (\ref{Vcsaba}).  
The thick curve shows the potential of Eq. (\ref{mmnsusyV}), 
which has a zero-energy bound state.  
\label{2vol}}
\end{center}
\end{figure}

What if a potential becomes positive at places but asymptotes to zero?  
This turns out to be of current interest in theories of higher
dimensions \cite{arkani}-\cite{dewolfe}.  
In these theories  one can be trying to discover if 
volcano-shaped  potentials have  zero-energy bound states
in what amounts to a 1-dimensional Schr\"odinger-equation \cite{mmnvolc}.   
An example would be the volcano potential  
\begin{equation}
V(z) = \frac{-\frac{1}{2} +\frac{19}{4}z^2}{[1+z^2]^2}.  
\label{Vcsaba}
\end{equation}
(See Figure \ref{2vol}.) 
But it can be demonstrated that the potential of Eq. \ref{Vcsaba} 
does not have a bound state by considering the potential 
( also shown in Figure \ref{2vol})
\begin{equation}
V(z) = \frac{-\left(\sqrt{5} - \frac{1}{2}\right) +\frac{19}{4}z^2}
           {[1+z^2]^2}.  
\label{mmnsusyV}
\end{equation} 
This is a supersymmetric potential \cite{mmns} of the form
\begin{eqnarray}
V(z) &=& \left[ W'(z)\right]^2 - W''(z).  \label {vasw} \\
\psi_0(z)&=&N \exp[-W(z)],  \label{psiasw} \\
W(z)&=&\left[\frac{\sqrt{5}}{2}-\frac{1}{4}\right] 
       \ln\left(1 + z^2\right).  \label{susyvl}
\end{eqnarray}   
Supersymmetry means  that the potential of 
Eq. (\ref{mmnsusyV}) must have a zero-energy bound state \cite{ze}.  
Therefore, since the potential of Eq. (\ref{mmnsusyV}) is everywhere 
below that of Eq. (\ref{Vcsaba}), the potential (\ref{Vcsaba}) can not 
have a  bound state \cite{df}.


There is also now the complementary idea that higher dimensions
might grow out of lower dimensions 
(instead of {\it vice versa}) \cite{georgi,hill}.  All these ideas 
stimulate the question
of just  what condition is necessary to have a bound state in
arbitrary {\it continuous} dimensions.  Here we will study  
this by using the analytically solvable square well potential.   
This may provide physical insight into the concept of how 
compactifying/expanding dimensions arise.  


\section{The D-dimensional Schr\"odinger equation}

We start with the integer $D$-dimensional, radial, Schr\"odinger equation 
\cite{louck}-\cite{mmnND2} 
\begin{equation}
E~R_D(z)=\left[-\frac{d^2}{dz^2} - \frac{(D-1)}{z}\frac{d}{dz} 
     + \frac{ l(l+D-2)}{z^2} + V(z)\right]R_D(z)~, 
\label{qmD}
\end{equation}
with volume element  $r^{D-1}dr$.
The angular momentum factor $l(l+D-2)$ comes from the D-dimensional 
spherical harmonics \cite{louck}. 
It is often useful to substitute
\begin{equation}
R_{D}(z)\equiv \frac{\chi_{D}(z)}{z^{(D-1)/2}} 
\label{ansatzd}
\end{equation}
into Eq. (\ref{qmD}).   This transforms the $D-$dimensional radial
Schr\"odinger equation  into an
effective $1-$dimensional Schr\"odinger equation
\begin{equation}
E \chi_{D}(z)=\left[-\frac{d^2}{dz^2}+ U_{D}(z)\right] \chi_{D}(z)~,
\label{qmDeff}
\end{equation}
where $U$ is an effective potential
\begin{eqnarray}
U_{D}(z)&=&\frac{(D-1)(D-3)}{4 z^2} + \frac{l(l+D-2)}{z^2}+
V(z)  \label{dork} \\
&=&
\frac{\left(l+\frac{D-3}{2}\right)  \left(l+\frac{D-1}{2}\right) }{z^2}+
V(z).    \label{dork2}
\end{eqnarray}

Now consider the radial bound-state problem in {\it continuous} 
dimensions $0 <D<\infty$ \cite{Dis0}.  To do this, we 
use the square well:
\begin{equation}
V(z) = \left\{ \begin{array}{ll} 
          -V_0, ~~~ & |z| < a, \\
                  & \\
          0, ~~~ & |z| > a. 
          \end{array} \right.
   \label{vwell}
\end{equation}
We are looking for the lowest energy eigenstate, so we can set $l=0$. 
Using the  notation 
\begin{equation}
v =V_0a^2,~~~~~~~~~~\epsilon = -E_0a^2 = |E_0|a^2, 
       ~~~~~~~~~ z = ya, 
\end{equation}
the internal and external Schr\"odinger equations are 
\begin{eqnarray}
0 & =& \left[-\frac{d^2}{dy^2} -\frac{D-1}{y}\frac{d}{dy}
-(v-\epsilon)\right] R_I , \label{scheqI} \\
0 & =& \left[-\frac{d^2}{dy^2} -\frac{D-1}{y}\frac{d}{dy}
+(\epsilon)\right] R_E \label{scheqE} .
\end{eqnarray}

The bound solutions to these equations are those that are finite at the
origin with zero slope there, are normalizable, and are valid for all 
$D>0$.  They are \cite{as}  
\begin{eqnarray}
R_I &\propto& 
  \frac{J_{(D-2)/2}(y\sqrt{v-\epsilon})}{(y\sqrt{v-\epsilon})^{(D-2)/2}} 
    \propto 
  \frac{j_{(D-3)/2}(y\sqrt{v-\epsilon})}{(y\sqrt{v-\epsilon})^{(D-3)/2}},
    \label{RI}    \\
R_E &\propto& 
\frac{K_{(D-2)/2}(y\sqrt{\epsilon})}{(y\sqrt{v\epsilon})^{(D-2)/2}} 
    \propto 
\frac{k_{(D-3)/2}(y\sqrt{\epsilon})}{(y\sqrt{\epsilon})^{(D-3)/2}},
    \label{RE}
\end{eqnarray}
where $J$ and $j$ are Bessel and spherical Bessel functions and 
$K$ and $k$ are modified Bessel and modified spherical Bessel functions. 

To obtain the  
bound, ground-state solution one needs to find the value of $\epsilon$
such that the internal ($y \le 1$ or $z\le a$) solutions,  
$R_I$, and external 
($y \ge 1$ or $z\ge a$) solutions,  $R_E$, satisfy the 
$y=1$  (or $z = a$) boundary condition 
\begin{equation}
\lim_{y\rightarrow 1}\left\{
      \frac{\frac{d}{dy}\left[R_I(y\sqrt{v-\epsilon})\right]}
               {R_I(y\sqrt{v-\epsilon})} \right\}
=  \lim_{y\rightarrow 1}\left\{
\frac{\frac{d}{dy}\left[R_E(y\sqrt{\epsilon})\right]}
               {R_E(y\sqrt{\epsilon})}\right\}.
\label{boundary}
\end{equation}
Eq. (\ref{boundary}) is equivalent to
\begin{equation}
\frac{\sqrt{v-\epsilon}J_{(D-2)/2}'(\sqrt{v-\epsilon})}
     {J_{(D-2)/2}(\sqrt{v-\epsilon})}
=\frac{\sqrt{\epsilon}K_{(D-2)/2}'(\sqrt{\epsilon})}
     {K_{(D-2)/2}(\sqrt{\epsilon})}
\label{boundary2}
\end{equation} 
where ``prime'' denotes derivative with respect to the argument.   
(Eq. (\ref{boundary}) can also be written in terms of the spherical Bessel
functions.)   Bessel-function recursion relations transform 
Eq. (\ref{boundary2}) to 
\begin{equation}
\frac{\sqrt{v-\epsilon}J_{D/2}(\sqrt{v-\epsilon})}
     {J_{(D-2)/2}(\sqrt{v-\epsilon})}
=\frac{\sqrt{\epsilon}K_{D/2}(\sqrt{\epsilon})}
     {K_{(D-2)/2}(\sqrt{\epsilon})}
\label{boundary3}
\end{equation} 


\section{Ground-state eigenenergies}

\subsection{Binding for dimensions $ \mathbf{0 < D < 2}$}

Here we are looking for the size of the binding energy 
$\epsilon_D(v)$ as 
$v(>\epsilon_D) \rightarrow 0$.  
We can use either analytic approximations 
or numerical methods to evaluate the 
Bessel functions in Eqs. 
(\ref{boundary})-(\ref{boundary3}).   To begin, one finds  that  
\begin{equation}
    \lim_{v\rightarrow0}
\epsilon_{D\rightarrow 0_+}  \sim  v_-. \label{b0}
\end{equation}
Examples are 
\begin{equation}
\epsilon_{0.01}(v=0.1) = 0.0986, ~~~~~~~~~~
\epsilon_{0.001}(v=0.1) = 0.0998. \label{b01}
\end{equation}
That is, as $D\rightarrow 0_+$ for small $v$, the particle becomes totally
bound, with  no quantum ground-state energy.   Actually,  for 
$D\rightarrow 0_+$  this tight binding holds for general well sizes,
e.g.,  
\begin{equation}
\epsilon_{0.01}(v=1) = 0.995, ~~~~~~~~~~
\epsilon_{0.001}(v=1) = 0.999. \label{b02}
\end{equation}

As $D$ increases, the binding becomes less tight.  When one 
reaches $D=1$ one has the standard 1-dimensional solution 
(which in spherical harmonic notation involves $j_{-1}$ and  $k_{-1}$)
\begin{equation}
\sqrt{\epsilon_1} 
= \sqrt{v-\epsilon_1}\tan\left(\sqrt{v-\epsilon_1}\right), 
~~~~~~~~~~
\epsilon_1 \sim v^2\left[1 - \frac{4}{3}v + \dots\right].  
\label{em1}
\end{equation}
In particular, 
\begin{equation}
\epsilon_{D=1}(v=0.1) = 0.0088. 
\label{d1bind}
\end{equation}

\begin{figure}[h]
 \begin{center}
\noindent    
\psfig{figure=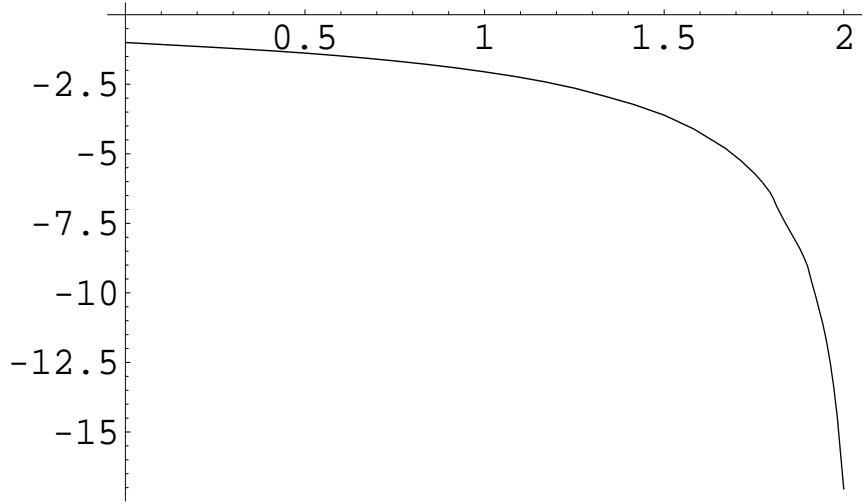,width=4.5in,height=3in}
\caption{$\log_{10}[\epsilon_D]$ is plotted as a function
of $D$ for $v= 0.1$  
\label{bd0to2}}
\end{center}
\end{figure}

As $0<D<2$ continues to rise, for a given small $v$
the binding energy gets smaller and smaller, varying as a power 
\begin{equation} 
\epsilon_D(v) \propto (v)^{\alpha(D)/(2-D)},
\end{equation}
where $\alpha$ is a decreasing function of D.  
(See Figure \ref{bd0to2}.)
Finally,  $\epsilon_D$ 
reaches a limit as $D \rightarrow 2_-$: 
\begin{equation}
\epsilon_{1.99999}(v=0.1) \approx 
\epsilon_{1.999999}(v=0.1) \approx 8.8 \times 10^{-18}.  \label{d0to2}
\end{equation}
This is 21 orders of magnitude smaller than the value 
of $\epsilon_{1}(v=0.1)$ for $D=1$ given in Eq. (\ref{d1bind}!  
(This limit will be explained in the next subsection.)  
But it is to be emphasized that the entire  restricted regime 
$0<2<D$ always has a bound state, no matter how small is $v$.  



\subsection{Binding for dimension $\mathbf{D=2}$}

When $D=2$, the interior and exterior wave functions are proportional to
the particularly elegant functions $J_0$ and $K_0$, which can be evaluated
especially easily for the  small arguments $\sqrt{v-\epsilon}$ 
and $\sqrt{\epsilon}$.  When these are used in Eq. (\ref{boundary}), one
finds
\begin{equation}
\lim_{v\rightarrow 0} \epsilon_{D=2}\sim \exp\left[-\frac{4}{(v+v^2/8)} 
       + 2(\ln 2 -\gamma)\right], 
\label{d2bindgen}
\end{equation}
where $\gamma$ is Euler's constant. 
Substituting in  $v=0.1$ gives  
\begin{equation}
\epsilon_{D=2}(v=0.1) = 8.8 \times 10^{-18},   
\label{d2bind}
\end{equation}
in agreement with our limiting result of Eq. (\ref{d0to2}). 

That is, the limit $D\rightarrow 2_-$ is continuous, and this limit 
yields an exponentially small binding.  This type of result for $D=2$  
is actually known \cite{textbook2}, 
although not as commonly as one might think.  
This result does not violate the 
general Rayleigh-Ritz principle \cite{2dvary}.   
One simply has to use something like an  exponentially
varying trial wave function of the form $\exp[-(r-r_0)^\alpha]$.
Further, this result has been given as an intuitive explanation of why
superconductivity works.  The small attraction still gives a bound state on
the 2-dimensional Fermi surface.


\subsection{Condition for binding when $\mathbf{D>2}$}

For $D>2$, there must always be a finite-sized $v>v_D$ for there to be a 
bound state. The way to determine $v_D$ is to solve the boundary equation 
(\ref{boundary}) for $\epsilon \rightarrow 0$. When this is done, 
the right-hand side reduces to $[-(D-2)]$ and the left-hand side reduces to 
$\left[-(D-2)/2 + \sqrt{v_D} 
\left(\left\{d/d\sqrt{v_D}\right\}J_{(D-2)/2}(\sqrt{v_D})\right)
/J_{(D-2)/2}(\sqrt{v_D})
\right]$.  
In other words, the problem amounts to finding the first zero of 
\begin{equation}
0= \left[\sqrt{v_{D}}~ J_D(\sqrt{v_{D}}) 
-(D-2) J_{(D-2)/2}(\sqrt{v_{D}})\right].  
\end{equation}

As $D\rightarrow 2_+$, one finds 
\begin{equation}
\lim_{\delta \rightarrow 0_+} v_{2+\delta} \sim 
4\left(\frac{D-2}{2}\right)= 2\delta.   \label{vdto2}
\end{equation}
For example, $v_{D=2.02} = 0.0402$

\begin{figure}[h]
 \begin{center}
\noindent    
\psfig{figure=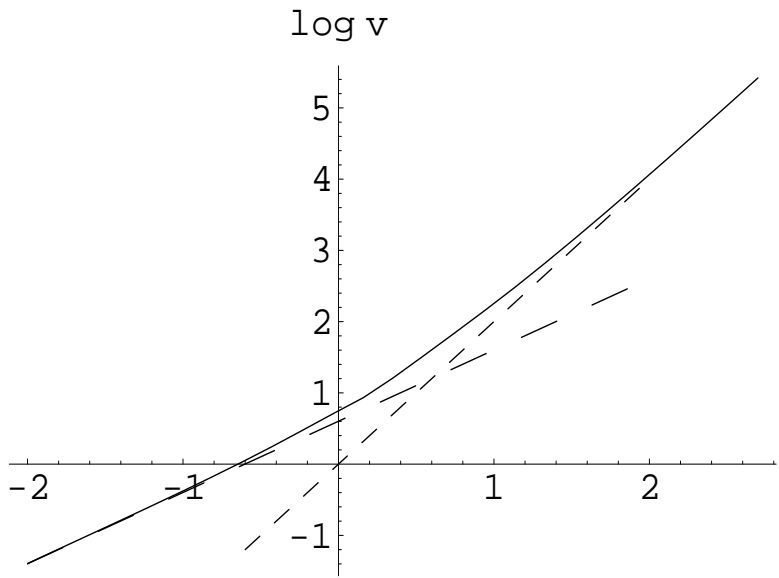,width=4.5in,height=3in}
\caption{A plot of $\log_{10}[v_D]$ as a function
of $\log_{10}\left[(D-2)^2/4\right]$.   The long-dash curve is the limit
curve of Eq. (\ref{vdto2}) and the short-dash curve is the limit curve of
Eq. (\ref{vdtoinf}). 
\label{bd2pl}}
\end{center}
\end{figure}

For odd-integer D, the boundary equation can be written in terms of 
integer-order spherical Bessel functions.  
These functions are in terms of powers and 
trigonometric functions. Although they become more complicated for higher
integers, the lower-order equations are simple:   
\begin{equation}
D=3:~~~~~~0 = \cot \sqrt{v_3}, ~~~~~~~~~~~~~~~~~~~~
  v>v_3=\frac{\pi^2}{4}\sim 2.5, 
\end{equation}
note that $\sqrt{v_3}=\pi/2$ is one unit of phase space, 
\begin{eqnarray}
D=5:~~~~~~0&=& \sin\sqrt{v_5} , ~~~~~~~~~~~~~~~~~~~
  v>v_5=4\frac{\pi^2}{4}\sim 9.9 \\
D=7:~~~~~~0&=& \tan \sqrt{v_7} -\sqrt{v_7}, ~~~~~~~~~
  v>v_7\sim 8.2~\frac{\pi^2}{4}\sim 20.
\end{eqnarray}
Thinking in units of $\nu = (D-2)/2$, one sees that $v_D$ is becoming
quadratic in this variable.  
Indeed, as $D$ becomes large, one finds  
\begin{equation}
\lim_{D\rightarrow\infty} v_D \sim  \left(\frac{D-2}{2}\right)^2. 
   \label{vdtoinf}
\end{equation}
An example is $v_{1000}\sim 2.63 \times 10^5$. 
The transition from the limit of Eq. (\ref{vdto2}) to the limit of Eq. 
(\ref{vdtoinf}) can be seen in Figure \ref{bd2pl}.
  

This all is consistent with the previous results; the higher the 
dimension the less is the binding.   It also agrees with known 
results in higher, integer $D=N$ dimensions.  For example, the ground-state
energies of the confining harmonic oscillator and the
infinitely deep hydrogen atom are (in ordinary units),
\begin{equation}
E_{0}^{HO}= \frac{N}{2}~ \hbar\omega, ~~~~~~~~~
E_{0}^{HA}= -\left(\frac{me^4}{2\hbar^2}\right)\frac{1}{[1+(N-3)/2]^2}, 
\end{equation}
respectively \cite{mmnND,HOHA}.


\section{Square-well volcanos}

Now we can return to the beginning and ask, ``Under what conditions will
one have a zero-energy bound (ground) state if one adds
an exterior angular-momentum barrier to the square well?''
\begin{equation}
V(z) = \left\{ \begin{array}{ll} 
          -V_0, ~~~ & |z| < a, \\
                  & \\
          +{b^2}/{z^2}, ~~~ & |z| > a. 
          \end{array} \right.
\end{equation}
The interior Schr\"odinger equation, its solution, and the boundary
condition will be the same as in Eqs. (\ref{scheqI}), (\ref{RI}), and 
(\ref{boundary})-(\ref{boundary3}), except that everywhere
$(v-\epsilon) \rightarrow v$.  The exterior Schr\"odinger equation is now 
\begin{equation}
0  = \left[-\frac{d^2}{dy^2} -\frac{D-1}{y}\frac{d}{dy}
+\frac{b^2}{y^2} \right] R_E.
\end{equation}
The solution and boundary condition are  
\begin{equation}
R_E(y) \propto y^{-s}, ~~~~~~~~~~~~~~~~   
\lim_{y\rightarrow 1}\left\{
\frac{\frac{d}{dy}\left[R_E(y)\right]}
               {R_E(y)}\right\} = -s,
\end{equation}
where 
\begin{equation}
b^2 = s(s+1) -s(D-1).  \label{b2}
\end{equation}
Normalizability of the wave function and positivity of Eq. (\ref{b2}) mean
\begin{equation}
S>\frac{D}{2}, ~~~~~~~~~~~~~~s>D-2.
\end{equation}

\begin{figure}[ht]
 \begin{center}
\noindent    
\psfig{figure=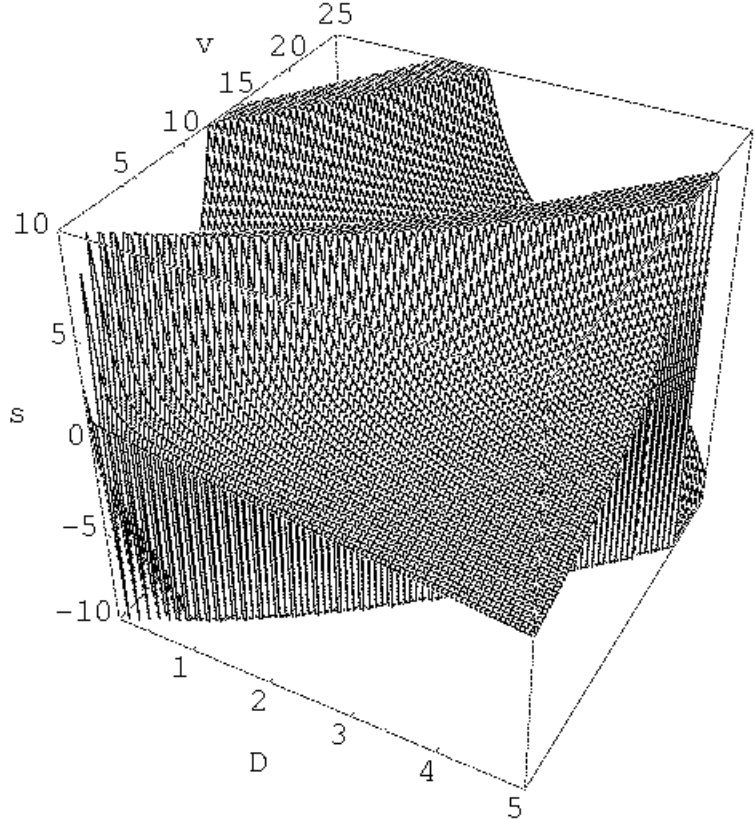,width=4.in,}
\caption{The solution of Eq. (\ref{newbound}), which is the condition  
for a zero-energy state.  The solution $s$  is plotted as a
function of $D$ and $v$.  [Finite numerical  cell size causes the narrow, 
infinite
peak/valley in $s$, near the $D-v$ origin, to appear finite in places.]
\label{Dvs}}
\end{center}
\end{figure}

Using these new boundary conditions one finds, analogously to
Eq. (\ref{boundary3}), 
that for the existence of a zero-energy bound state
one needs 
\begin{equation}
s = \frac{\sqrt{v}J_{D/2}(\sqrt{v})}{J_{(D-2)/2}(\sqrt{v}) }    
- \frac{(D-2)}{2}.  \label{newbound}
\end{equation}
Figure \ref{Dvs} shows the solution for $s$ as a function of $D\le 5$ and
$v \le 25 $.  For low $\{D,v\}$ there is a widening infinite ridge as one
goes to higher $v>D$, followed
by an infinite valley.  For higher $v$ a new (wider) infinite ridge/valley
sequence begins.  (Further sequences, not shown, start for yet higher $v$.) 


\begin{figure}[h!] 
    \noindent
    \begin{center}  
        \begin{minipage}[t]{.46\linewidth} 
            \epsfig{file=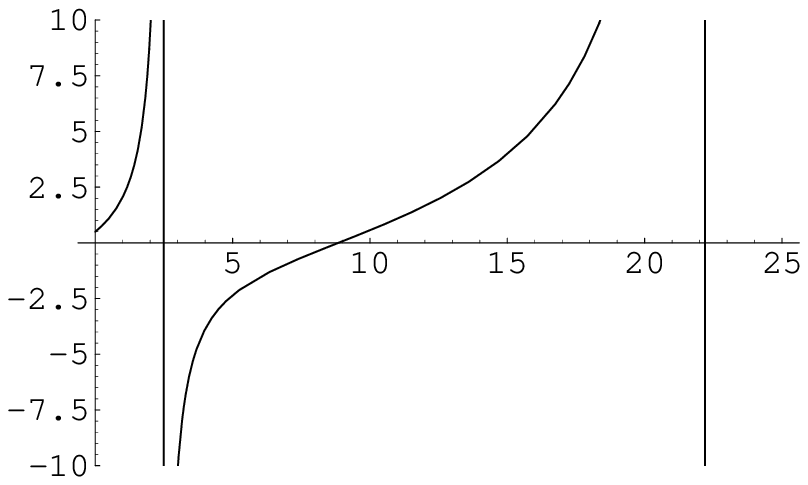,width=76mm}
            \noindent
        \end{minipage}
        \hskip 15pt
        \begin{minipage}[t]{.46\linewidth}
            \epsfig{file=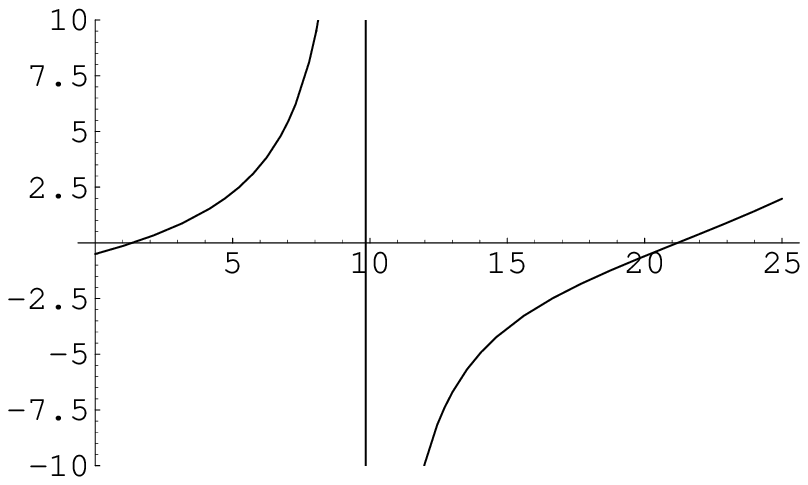,width=76mm}
            \noindent
        \end{minipage}
\caption{Plots of $s$ vs. $v$ for the cases $D=1$ and $D=3$. }
            \label{2slice}  
    \end{center}
\end{figure}

To better understand this result, in Figure \ref{2slice} 
we show two physically illuminating $D$ slices,   
$D=1$ and $D=3$.   For $D=1$ one needs $s>1/2$, so that the zero 
energy state is normalizable and hence a bound ``ground state.'' 
This is the asymptotic result as $v \rightarrow 0$.  As $v$
gets larger the height of the tail must get larger to force the ground
state to remain at zero energy.  Finally, at the depth of the ground state for
an infinite well, $v=\pi^2/4$, the tail is infinitely strong, making it an 
effective infinite well.  For still larger $v$ the solution becomes 
discontinuously unphysical, rising from 
negative infinity.  The solution eventually becomes positive (but not
bound) since at first $s<1/2$.  When $s>1/2$ one has a new bound solution,
which is  an excited state.  This can be verified when the
previous pattern is repeated (with a wider
peak).  There is a second discontinuous peak/valley jump at $v=\pi^2/4$,   
the energy of the first (even) excited state for the
infinite square well.  Finally, as expected, the $D=3$ slice shows no
bound-state zero-energy solution for small $v$.  It is only when $s>1/2$
that a solution exists.  The wider peak/valley discontinuity 
is at the energy of the 3-dimensional infinite-well ground state,$v=\pi^2$. 

\section{Discussion}

If one goes back to Eq. (\ref{dork2}) one sees that for $l=0$ it is  
the dimensions $1 < D< 3$ that have an ``effective attractive'' 
angular-momentum barrier for the $\chi$ equation 
(\ref{qmDeff}).  Among this set of dimensions,  $D=2$ is 
the only integer dimension  \cite{wolf2d}.  
The other $D$ with this property have power-law binding ($1<D<2$) 
or need a finite $v_D$ to have binding ($2<D<3$).  
$D=2$  is truly the boundary case for quantum binding \cite{wet}.  
Further, $1<D<3$ is a transition region in general.  If one looks at
Figure \ref{bd0to2}, one sees that it is for  $D>1$ that the binding energy
starts to go steeply towards zero.  
Also, if one looks at Figure \ref{bd2pl}, it is for $D>3$ 
[or $\log_{10}\left\{(D-2)^2/4\right\}>(-0.60)$ 
that $v_D$ goes over to the more rapidly rising  quadratic function of $D$.  

It is to be noted that how one handles the spherical harmonics in 
continuous dimensions is a separate interesting question.  Specifically, 
the boundary conditions will be related to questions of statistics, 
as in the fractional quantum Hall effect.  One also can consider the 
introduction of spin and/or explicit relativistic binding.  
These are topics for further investigation. 

Here the energetics of bound states in continuous radial dimensions has 
been investigated.  This could shed light on the energetics of
changing dimensions in the physical universe.


\section*{Acknowledgements} 

I first thank Csaba Cs\'aki Joshua Erlich, who interested me in the 
bound-state problem as it relates to 
fundamental theories of extra dimensions.
Wolfgang Schleich raised the question about the odd behavior of
the 2-dimensional bound-state problem.  It was these two stimuli that
initiated this work.  I also am very grateful for the comments and
suggestions of Kurt Gottfried, 
Andr\'e Martin, Eugen Merzbacher, and Jean-Marc Richard.
This work was supported by the US DOE.    




\begin{thebibliography}{99}
\bibitem{ze}  Since the potential goes to infinity as $+1/z^\nu$, 
$\nu > 1/2$, there is already a bound state at zero energy, instead of 
only at infinitesimally small negative energy.
\bibitem{Dis0}  The limit $D=0$ would yield an interior solution that is not
normalizable.
\end{thebibliography}
\end{document}